\documentclass[conference]{IEEEtran}
\IEEEoverridecommandlockouts
\usepackage{cite}
\usepackage{amsmath,amssymb,amsfonts}
\usepackage{algorithmic}
\usepackage{graphicx}
\usepackage{siunitx}
\usepackage{tipa}
\usepackage{textcomp}
\usepackage{xcolor}
\usepackage{multirow}
\usepackage{booktabs}
\usepackage[hidelinks]{hyperref}
\def\BibTeX{{\rm B\kern-.05em{\sc i\kern-.025em b}\kern-.08em
    T\kern-.1667em\lower.7ex\hbox{E}\kern-.125emX}}
\begin{document}

\title{Evaluating Multichannel Speech Enhancement Algorithms at the Phoneme Scale Across Genders
}

\author{
    \IEEEauthorblockN{Nasser-Eddine Monir, Paul Magron, Romain Serizel}
    \IEEEauthorblockA{\textit{Université de Lorraine, CNRS, Inria, LORIA} \\
    F-54000 Nancy, France \\
    \{nasser-eddine.monir, paul.magron\}@inria.fr, romain.serizel@loria.fr}
}

\maketitle

\begin{abstract}
Multichannel speech enhancement algorithms are essential for improving the intelligibility of speech signals in noisy environments. These algorithms are usually evaluated at the utterance level, but this approach overlooks the disparities in acoustic characteristics that are observed in different phoneme categories and between male and female speakers. In this paper, we investigate the impact of gender and phonetic content on speech enhancement algorithms. We motivate this approach by outlining phoneme- and gender-specific spectral features.
Our experiments reveal that while utterance-level differences between genders are minimal, significant variations emerge at the phoneme level.
Results show that the tested algorithms better reduce interference with fewer artifacts on female speech, particularly in plosives, fricatives, and vowels. Additionally, they demonstrate greater performance for female speech in terms of perceptual and speech recognition metrics.
\end{abstract}

\section{Introduction}
\label{sec:intro}

Speech enhancement (SE) aims at retrieving a clean speech signal from a mixture contaminated with noise and/or reverberation. SE finds application in many downstream tasks such as hearing aids~\cite{Esra2024}, speech recognition~\cite{iwamoto2022bad}, and audio conferencing~\cite{rao2021interspeech}.
%While speech enhancement algorithms broadly encompass noise reduction and dereverberation, this article focuses on the former.
Traditional SE algorithms rely on signal processing techniques, leveraging mathematical assumptions about speech and noise~\cite{Benesty2008SpringerHandbook, Hendriks2011}. However, modern approaches are data-driven, and predominantly use deep neural networks (DNNs). A common strategy consists in combining DNNs for estimating spectral parameters (e.g., a correlation matrix or a time-frequency spectrum) with a traditional spatial filter, such as a minimum variance distortionless beamformer (MVDR) or a multichannel Wiener filter~\cite{Heymann2016,Nugraha2016,Coto-Jimenez2018,Liu2018mvdr,Carbajal2020,Furnon2021}. Alternatively, some algorithms directly estimate enhanced signals or filters via DNNs~\cite{wang18h_interspeech, Luo2019}.%, Tolooshams2020, ren21_interspeech}.

SE is typically evaluated at the utterance level using metrics such as signal-to-distortion, artifacts, or interference ratios (SDR, SIR, SAR)~\cite{Vincent2006bsseval,LeRoux2019sisdr}.
%, or their scale-invariant counterparts~\cite{LeRoux2019sisdr}. 
However, Miller et al.~\cite{MillerNicely1955} emphasize the variability in phoneme noise tolerance, highlighting the importance of a nuanced understanding of how phonemes, particularly consonants and vowels, are affected by noise. Adachi et al.~\cite{adachi2006intelligibility} reveal the ways in which different phonemes are impacted by noise for both native and non-native speakers, while Meyer et al.~\cite{meyer2010human, zaar2017predicting} observe confusion among phonemes within both human perception and automatic speech recognition frameworks, indicating that consonants and vowels are differently affected by the loss of information due to noise exposure. This suggests that evaluating SE using utterance-level metrics may overlook the detailed impacts of noise on different phonemes and the algorithms' processing of these sounds. This has motivated us to evaluate SE algorithms at the phoneme scale in a previous study~\cite{monir2024phonemescale}.
%Building on our previous work~\cite{monir2024phonemescale}, which assessed multichannel speech enhancement algorithms at the phoneme scale, this study extends the analysis by investigating whether algorithms differentiate male and female speech in terms of phoneme-level performance.

However, beyond overall variability in phonemes, significant acoustic variations between male and female voices reveal that there are gender-specific differences in phonemes~\cite{Henton1986}. Gender perception in voices is mainly related to the fundamental frequency that is due to the length of the vocal tract, which affects formant patterns~\cite{coleman1976comparison,pepiot}. Calliope~\cite{calliope1989parole} examined gender-based differences in vocalic formants, influencing the performance of data-driven speech processing~\cite{addadecker05_interspeech}. These studies highlight the need to consider phonetic gender variations in SE technologies.

In this paper, we extend our previous work~\cite{monir2024phonemescale} by investigating the impact of gender and phonetic content on SE algorithms. We motivate our approach by analyzing spectral disparities in phonemes and gender. We evaluate three state-of-the-art multichannel SE algorithms~\cite{Furnon2021,Fasnet, Heymann2016} in a realistic simulated acoustic scenario using various metrics. The results reveal that while overall enhancement performance at the utterance level shows minimal gender differences, a deeper analysis at the phoneme level uncovers distinct trends, with female speech often exhibiting greater interference reduction and perceptual quality improvements, particularly as noise levels decrease. 
% \textcolor{red}{The rest of this paper is structured as follows. Section~\ref{sec:methodo} details our evaluation methodology. Section~\ref{sec:exp} details the experimental setup, and Section~\ref{sec:results} discusses the results. Finally, Section~\ref{sec:conclusion} concludes this paper.}

\section{Methodology}
\label{sec:methodo}
In this section, we describe our methodology for analyzing the impact of SE algorithms on male and female speech at a phoneme level.

\subsection{Evaluation at the phoneme level}

\begin{figure}[t]
  \centering
  \includegraphics[width=1.0\linewidth]{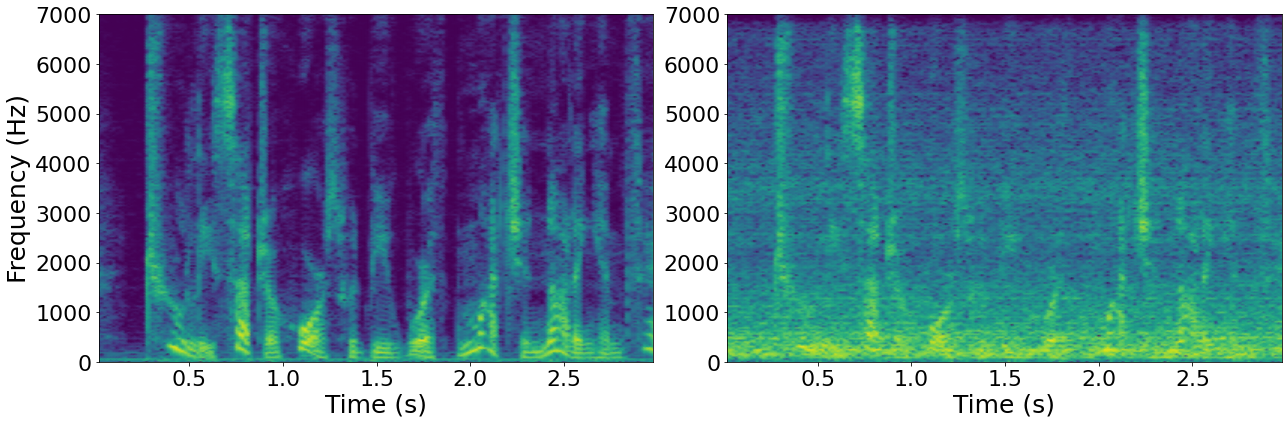}
  \caption{Spectrograms of a male speech signal: clean (left) and mixed with an SSN at -5~dB SNR (right).}
  \label{fig:spectrograms}
\end{figure}

SE algorithms are typically evaluated at the utterance level, which provides an overall measure of speech clarity and comprehension. However, Miller et al.~\cite{MillerNicely1955} suggest that consonants and vowels are impacted to different extents by the loss of information caused by the presence of noise.

To illustrate this, we display in Figure~\ref{fig:spectrograms} the spectrogram of a clean speech signal and its mixture with a speech-shaped noise (SSN, see Section~\ref{ssec:data}). We observe that the low-frequency content of the clean speech is masked when mixed with the noise, while only the sharp bursts at high frequencies remain slightly visible. Such high frequency components are indicative of phonemes that are characterized by their wide-band acoustic content, such as plosives or fricatives. This motivates evaluating speech across phonemes rather than solely at the utterance level, as initiated in our previous work~\cite{monir2024phonemescale}.

\subsection{Evaluation depending on the gender of the speaker}

\begin{figure}[t]
  \centering
  \includegraphics[width=1\columnwidth]{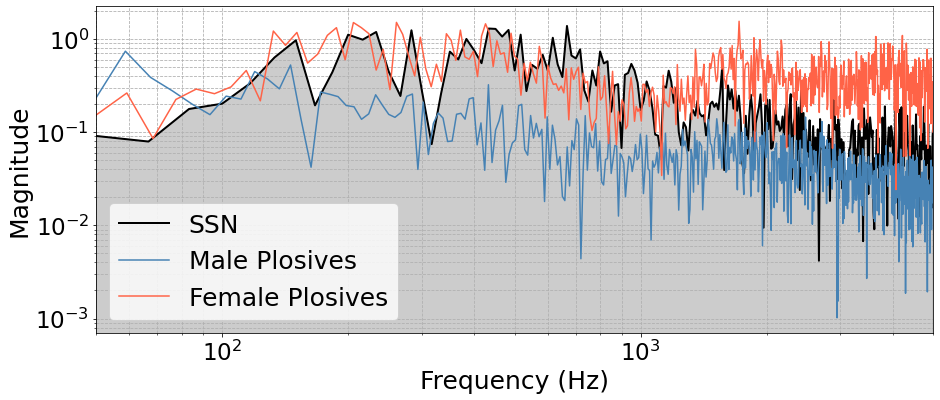}
  \caption{Spectrum of the noise and clean plosives for male and female speakers at 0~dB SNR, computed from the dry signals.}
  \label{fig:plosives_spectra_fm}
\end{figure}

Beyond phonemic variability, speakers of different genders can introduce differences in speech acoustics. We illustrate this phenomenon by displaying a range of plosive sounds from male and female speakers in Figure~\ref{fig:plosives_spectra_fm}, and we analyze how male and female plosives respond to SSN, highlighting spectral overlap and differential noise masking effects across genders. This invites a closer analysis of how speech enhancement algorithms might optimally address these gender-specific phonetic characteristics.

Male plosives show strong intensity below $100$~Hz, while female plosives dominate at frequencies above $100$~Hz. This indicates a shift in spectral emphasis, with male speech contributing more to the low-frequency range and female speech being more prominent in the mid to high frequencies. Both male and female plosives are partially masked by the SSN, but female plosives maintain stronger magnitudes above $100$~Hz. These differences also appear in near-close vowels and fricatives at low noise levels, inviting further investigation into male-female speech characteristics to better understand their processing by SE algorithms.

\section{Experimental Setup}
\label{sec:exp}

In this section we detail our experimental protocol. For a reproducibility purpose, both our code and the pretrained model weights are available online\footnote{\href{https://github.com/Nasseredd/mcse-phg}{https://github.com/Nasseredd/mcse-phg}}.

\subsection{Acoustic scenarios and dataset}
\label{ssec:data}

We build a dataset from LibriSpeech~\cite{Panayotov2015librispeech}, using the \texttt{train-clean-100}, \texttt{dev-clean}, and \texttt{test-clean} subsets for training, validation, and testing, ensuring balanced male and female durations. For each subset, only 50\% of the data is used as clean speech, yielding 50h for training and approximately 2.5h each for validation and testing. The other 50\% are used to generate SSN, contributing 30\% of total noise. SSN was chosen to provide controlled experimental conditions while preserving the spectral characteristics of both male and female voices. The remaining 70\% comes from ecological sources in Disco-noise~\cite{Furnon2021}. The validation set follows the same process, while the test set uses only SSN. SSN is generated by computing the discrete Fourier transform (DFT) of five male and five female speech signal, randomizing its phase, and applying the inverse DFT to ensure spectral consistency. %All speech is standardized to 10 seconds.

We simulate a hearing aid setup with four microphones, two on each ear. For training and validation, room impulse responses (RIRs) are generated using Pyroomacoustics~\cite{Scheibler2018} with $RT_{60}$ between 0.15--0.4~s, and room dimensions of 3--8~m (length), 3--5 m (width), and 2.5--3~m (height). Speech and noise sources are randomly positioned, with signal-to-noise ratios (SNRs) from -10~dB to 10~dB, computed on the dry signals. For testing, we use measured RIRs~\cite{binaurec} placing the speech source at 0~degrees (directly ahead of the listener) and the noise at 45~degrees to the right of the listener at SNR levels of -5, 0, or 5~dB. % This setup produces four input channels (two per ear) and one output channel per ear.

\subsection{Phoneme segmentation}

We use the Montreal Forced Aligner (MFA)~\cite{McAuliffe2017MontrealFA} to segment speech into phonemes, according to the international phonetic alphabet (IPA) chart in MFA. The English MFA dictionary v2.2.1 includes 13 phoneme classes (8 consonants and 5 vowels), and we adopt an extended classification from Monir et al.~\cite{monir2024phonemescale}, adding a vowel class for near-close phonemes \textipa{/[I]/} and \textipa{/[U]/}.

\subsection{Speech enhancement algorithms}

We perform SE with three algorithms. Tango~\cite{Furnon2021} is a hybrid algorithm derived from the DANSE algorithm~\cite{5483076}. It employs a convolutional recurrent neural network for estimating time-frequency (TF) masks with binaural cues. FaSNet~\cite{Fasnet} is an end-to-end time-domain beamformer. It processes time-domain features with a dual-path recurrent network to estimate spectral masks for beamforming. MVDR~\cite{Heymann2016} is a frequency-domain beamformer technique that uses a bidirectional long short-term memory network to predict TF masks that are used to estimate speech and noise covariance matrices.
Note that our implementation (including training and evaluation scripts) rely on the Asteroid~\cite{Pariente2020Asteroid} and ESPnet~\cite{watanabe2018espnet} toolboxes.
Full training details (e.g., loss functions, optimizers, etc.) are available in our code. Finally, as this study does not aim to compare algorithms, we average their results to focus on gender and phoneme differences.

\subsection{Evaluation metrics}

\begin{figure}[t]
  \centering
  \includegraphics[width=1.0 \columnwidth]{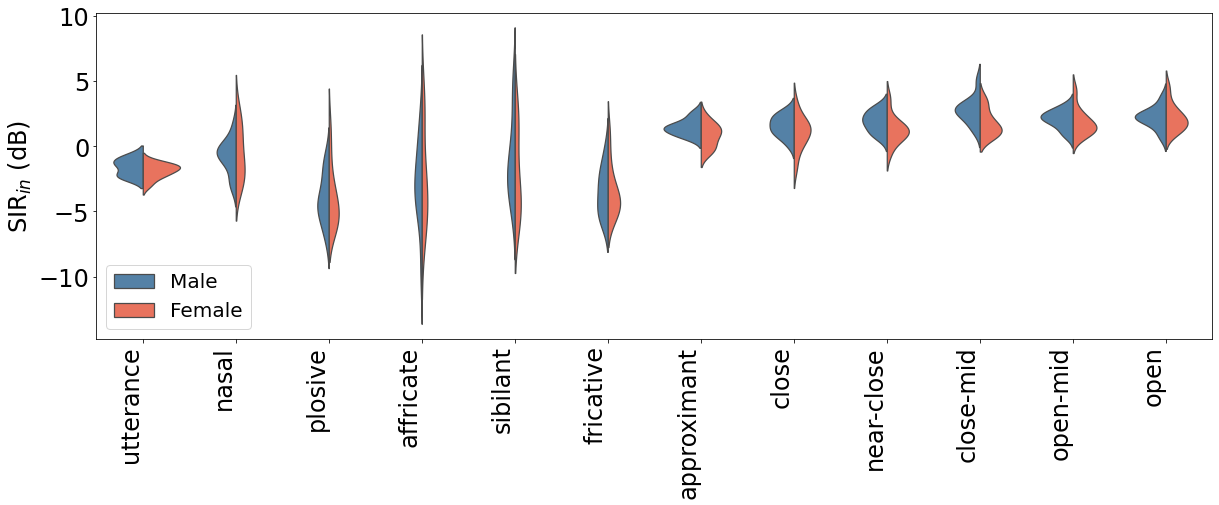}
  \caption{Input SIR at the utterance level and across phoneme categories and genders at 0~dB SNR. Each violin plot represents the approximate distribution of the data.}
  \label{fig:figure_sir_in_0dB}
\end{figure}

\begin{table}[t]
\centering
\caption{Mean SIR$_{in}$ Across Genders and SNR Levels.}
\label{tab:metrics_gender_sir_in}
\small  % Reduce font size for better fit in IEEEtran
\begin{tabular}{l cccccc}
\hline
 & \multicolumn{2}{c}{-5~dB} & \multicolumn{2}{c}{0~dB} & \multicolumn{2}{c}{5~dB} \\
\cline{2-3} \cline{4-5} \cline{6-7}
        & M        & F      & M        & F      & M        & F \\
\hline
Consonants & -7.71   & -7.87     & -1.16     & -1.45     & 4.49      & 4.07 \\
Vowels     & -3.69   & -3.79     & 2.10      & 1.90      & 7.46      & 7.19 \\
\hline
\end{tabular}
\vspace{0.1cm}  % Add a small vertical space
\scriptsize  % Use a smaller font size for the note
\\\textit{Note:} M = Male, F = Female.
\end{table}

% P-VALUES SIR_in
% \begin{table}[h]
% \centering
% \caption{Mann-Whitney U Test Results ($p$-value) for SIR$_{in}$.}
% \label{tab:cv_sir_in}
% \begin{tabular}{l c c c}
% \hline
%  & -5~dB & 0~dB & 5~dB \\
% %Comparison & $p$-value & $p$-value & $p$-value \\
% \hline
% Consonants vs. Vowels (Male)   & \textbf{0.0002} & \textbf{0.0003} & \textbf{0.0008} \\
% Consonants vs. Vowels (Female) & \textbf{0.0002} & \textbf{0.0017} & \textbf{0.0013} \\
% Male vs. Female (Consonants)   & 1.0000  & 0.7913  & 0.5205  \\
% Male vs. Female (Vowels)       & 0.5708  & 0.1041  & 0.1859  \\
% \hline
% \end{tabular}
% \end{table}

We evaluate
SE using the scale-invariant~\cite{LeRoux2019sisdr} SIR and SAR, expressed in~dB~\cite{Vincent2006bsseval}. We report the input and output SIR, respectively denoted SIR$_{in}$ and SIR$_{out}$. SIR$_{in}$ is the SIR at the ear level, thus it accounts for room acoustics (as opposed to the SNR which is adjusted on dry sources). SIR$_{out}$ measures the residual interference after performing SE. Similarly, the output SAR (denoted SAR$_{out}$) assesses the overall amount of artifacts after SE. We do not report the input SAR as it is theoretically infinite when no processing has been applied.

Additionally, we report the perceptual evaluation of speech quality (PESQ)~\cite{Rix2001}, short-time objective intelligibility (STOI)~\cite{STOI}, and hearing aid speech perception index\footnote{We use the HASPI-v2 version with the normal-hearing auditory model.} (HASPI)~\cite{haspi} scores. To assess improvements in perceived quality and intelligibility, we measure PESQ and STOI before (PESQ$_{in}$, STOI$_{in}$) and after enhancement (PESQ$_{out}$, STOI$_{out}$), as well as their difference $\Delta_{\text{PESQ}}$ and $\Delta_{\text{STOI}}$.
%HASPI evaluates speech perception in hearing aids.

Finally, we feed the enhanced speech to five automatic speech recognition models\footnote{Except for Whisper, all models are pre-trained on the gender-balanced LibriSpeech dataset.} (Wav2vec, Wav2vec-lv60, Conformer-CTC, Conformer-Transducer and Whisper~\cite{wav2vec2.0,conformer,whisper}), selected as proxies for evaluating speech intelligibility. We then compute the average word error rate (WER) over models, which is an overall measure of the impact of SE on speech recognition performance.

To assess statistical significance, we conduct Mann-Whitney U tests, a non-parametric method that is suited for comparing independent samples drawn from two unknown distributions of any of the afore-mentioned metrics.
%either SIR$_{in}$, SIR$_{out}$, or SAR$_{out}$.
The statistical tests primarily compare male and female speech across different levels: at the utterance level, within consonants and vowels separately, and for each phoneme category. The method computes a $p$-value for a given pair of input distributions, and we consider the difference between categories to be significant when ${p<0.05}$. While all statistical analyses were conducted, $p$-values are not systematically reported to ensure readability and focus on the most relevant findings, which are discussed in the text.

\section{Results and Discussion}
\label{sec:results}

\subsection{Analysis on input signals}

\begin{figure}[t]
  \centering
  \includegraphics[width=1.0 \columnwidth]{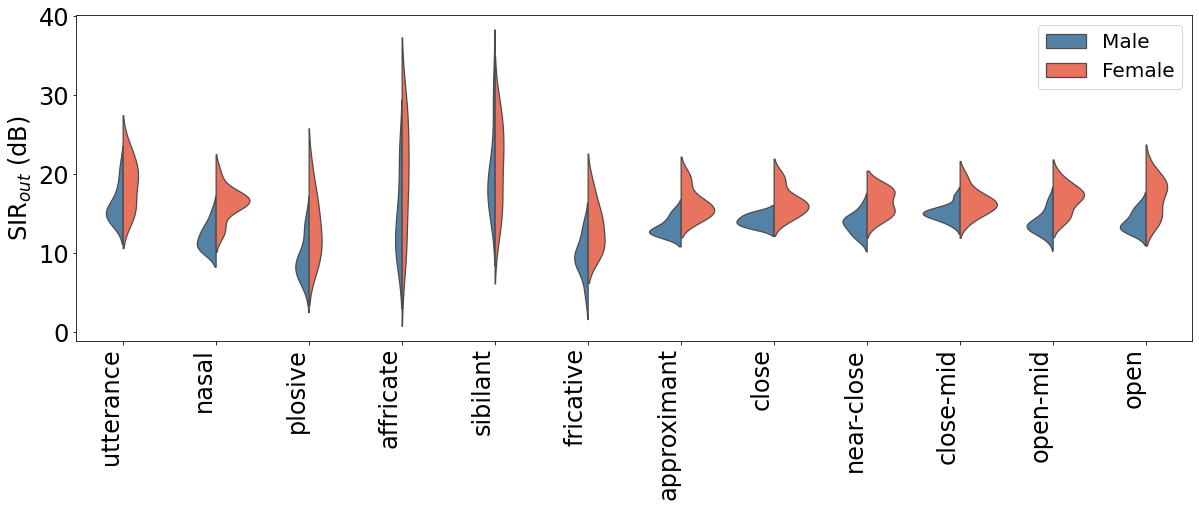}
  \caption{Output SIR at the utterance level and across phoneme categories and genders at 0~dB SNR.}
  \label{fig:figure_sir_out_0dB}
\end{figure}

\begin{table}[t]
\centering
\caption{Mean SIR$_{out}$ Across Genders and SNR Levels.}
\label{tab:metrics_gender_sir_out}
\small  % Reduce font size for better fit in IEEEtran
\begin{tabular}{l cccccc}
\hline
 & \multicolumn{2}{c}{-5~dB} & \multicolumn{2}{c}{0~dB} & \multicolumn{2}{c}{5~dB} \\
\cline{2-3} \cline{4-5} \cline{6-7}
% \vspace{0.1cm}  % Add a small vertical space
        & M        & F      & M        & F      & M        & F \\
\hline
Consonants & 7.87   & 8.13     & 14.38     & 17.18     & 20.11      & 23.35 \\
Vowels     & 8.50   & 11.45     & 14.08      & 16.97      & 18.88      & 21.52 \\
\hline
\end{tabular}
% \vspace{0.1cm}  % Add a small vertical space
% \scriptsize  % Use a smaller font size for the note
% \textit{Note:} M = Male, F = Female.
\end{table}

% P-VALUE SIR_out
% \begin{table}[t]
% \centering
% \caption{Mann-Whitney U Test Results ($p$-value) for SIR$_{out}$.}
% \begin{tabular}{l c c c}
% \hline
%  & -5~dB & 0~dB & 5~dB \\
% \hline
% Male vs. Female Consonants & 0.0640 & \textbf{0.0173} & \textbf{0.0257} \\
% Male vs. Female Vowels & \textbf{0.0017} & \textbf{0.0022} & \textbf{0.0036} \\
% \hline
% \end{tabular}
% \label{tab:cv_sir_out}
% \end{table}

First, we analyze the input signals before performing SE. The results in terms of SIR$_{in}$ over phoneme categories and genders at 0~dB SNR\footnote{Similar trends can be observed at -5 and 5~dB, but figures are omitted due to space constraints. This also applies to Figures~\ref{fig:figure_sir_out_0dB} and~\ref{fig:figure_sar_out_0dB}.} are displayed in
Figure~\ref{fig:figure_sir_in_0dB}.
We observe that interfering noise at the utterance level is similar for males and females, and no significant difference between gender can be observed within each phoneme category.
%Figure~\ref{fig:figure_sir_in_0dB} shows that interfering noise at the utterance level is similar for males and females at 0~dB SNR, a trend also observed at -5~dB and 5~dB\footnote{Figures omitted due to space constraints.}.

Table~\ref{tab:metrics_gender_sir_in} presents the mean SIR$_{in}$ values across SNR levels for male and female speakers. The results show statistically significant differences between consonants and vowels for both male ($p<0.001$) and female speakers ($p<0.002$) across SNR levels.
%($p=0.0002$, $p=0.0003$, and $p=0.0008$ at -5, 0, and 5~dB, respectively) and female speakers ($p=0.0002$, $p=0.0017$, and $p=0.0013$ at -5, 0, and 5~dB, respectively). 
However, no significant differences between genders were found within each phoneme category at any SNR level. This gender similarity extends across phoneme categories, where no significant difference is observed, except for underrepresented laterals and taps. This suggests that speech sounds has a greater influence on SIR$_{in}$ than genders.

%influences noise interference more than gender, indicating that male and female voices experience comparable noise effects within this enclosed room.

\subsection{Results after speech enhancement}

%We now examine the impact of SE across phonemes and genders.%, thus in terms of output metrics.

\subsubsection{Impact on interference}

% We assess the performance of speech enhancement (SE) algorithms by examining SIR$_{out}$ across phoneme categories and genders, as shown in Figure~\ref{fig:figure_sir_out_0dB} at 0~dB SNR.

First, we analyze the impact of SE in terms of noise reduction, as measured by the output SIR. We display in Figure~\ref{fig:figure_sir_out_0dB} the output SIR at 0~dB across utterance, phoneme categories and genders, while Table~\ref{tab:metrics_gender_sir_out} summarizes the mean SIR$_{out}$ for consonants and vowels by gender across SNR levels. We observe in Figure~\ref{fig:figure_sir_out_0dB} that SE algorithms process male and female speech similarly at the utterance level (${p=0.07}$).
However, results in Table~\ref{tab:metrics_gender_sir_out} show significant performance differences at 5~dB and 0~dB for consonants ($p=0.025$ and $0.017$) and vowels ($p=0.002$ and $0.003$). This suggests that the algorithms process interference differently depending on gender-specific speech characteristics. Nonetheless, at -5~dB, the difference for consonants disappears ($p=0.064$), likely because the interfering noise masks differences in spectral and temporal cues.
%likely because the interfering noise masks subtle spectral and temporal cues that differentiate.
Despite vowels also being susceptible to noise masking, we observe gender differences ($p=0.001$) at the output.
% In contrast, vowels maintain significant differences, possibly due to their longer duration and more stable formant structure, making them less susceptible to noise masking.

While the overall enhancement performance appears similar across genders when averaged over entire utterances, a closer look at individual phoneme categories in Figure~\ref{fig:figure_sir_out_0dB} reveals significant differences in nasals, plosives, fricatives, approximants, and most vowel types. This suggests that male and female speakers exhibit distinct acoustic properties in these phonemes, which persist even after interference reduction. On the other hand, affricates and sibilants show no significant gender-based difference ($p=0.09$ and $0.79$, respectively), likely because these phonemes naturally contain turbulent, high-frequency energy, making them harder to distinguish from background noise.

\subsubsection{Impact on artifacts}

\begin{figure}[t]
  \centering
  \includegraphics[width=1.0 \columnwidth]{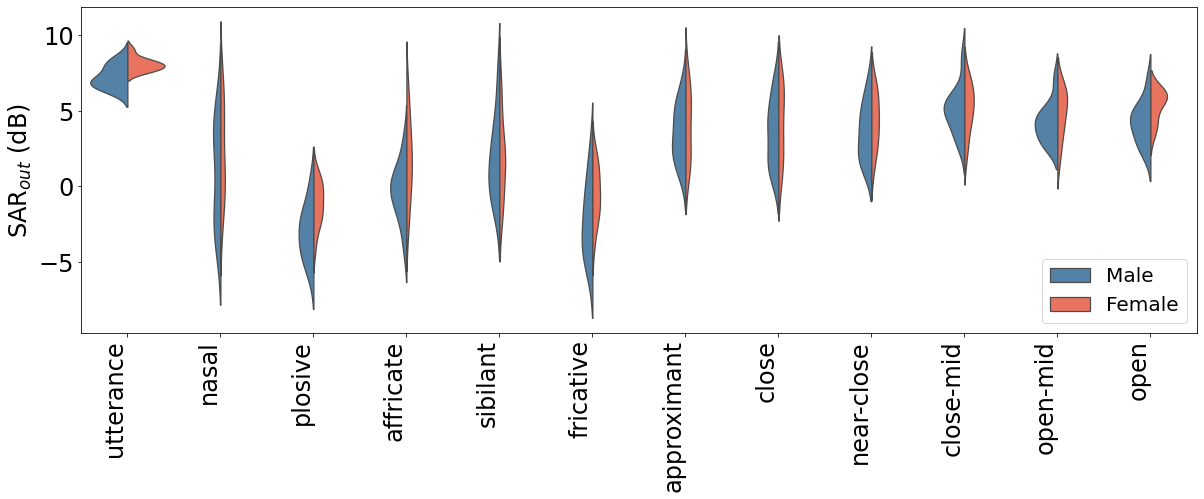}
  \caption{Output SAR at the utterance level and across phoneme categories and genders at 0~dB SNR.}
  \label{fig:figure_sar_out_0dB}
\end{figure}

\begin{table}[t]
\centering
\caption{Mean SAR$_{out}$ Across Genders and SNR Levels.}
\label{tab:metrics_gender_sar_out}
\small  % Reduce font size for better fit in IEEEtran
\begin{tabular}{l cccccc}
\hline
 & \multicolumn{2}{c}{-5~dB} & \multicolumn{2}{c}{0~dB} & \multicolumn{2}{c}{5~dB} \\
\cline{2-3} \cline{4-5} \cline{6-7}
        & M        & F      & M        & F      & M        & F \\
\hline
Consonants & -1.33   & -1.11     & 0.89     & 1.90     & 2.03      & 3.99 \\
Vowels     & 1.21   & 1.69     & 4.12      & 5.01      & 5.56      & 7.18 \\
\hline
\end{tabular}
% \vspace{0.1cm}  % Add a small vertical space
% \scriptsize  % Use a smaller font size for the note
% \textit{Note:} M = Male, F = Female.
\end{table}

Here we analyze the impact of SE in terms of artifacts in the estimated signals. Figure~\ref{fig:figure_sar_out_0dB} displays the output SAR across phoneme categories and genders at 0~dB. The results suggest that at the utterance level, SE algorithms preserve female speech quality slightly better than male speech ($p=0.02$). This trend remains at 5~dB, but the difference between genders becomes more pronounced at -5~dB.

Table~\ref{tab:metrics_gender_sar_out} presents the mean SAR$_{out}$ for consonants and vowels by gender across SNR levels. The results show no significant gender differences for consonants ($p=0.24$) and vowels ($p=0.18$) at 0~dB, or other tested SNR levels. This indicates that, on average, the algorithm does not introduce artifacts in a gender-biased way when considering broad phoneme categories.

At the phoneme level, however, more nuanced differences emerge. Plosives show a significant difference ($p=0.02$), with female speech having a higher output SAR, which means that male plosives are more affected by artifacts. This suggests that the algorithms show limitations in preserving plosive sounds in male speech, possibly due to their stronger bursts and lower fundamental frequencies. In contrast, nasals, affricates, fricatives, approximants, and vowels do not show statistically significant differences, suggesting that the algorithm affects these sounds similarly across genders.

\subsubsection{Impact on speech recognition}

%\begin{figure}[t]
%  \centering
%  \includegraphics[width=1.0 \columnwidth]{img/wer_est.png}
%  \caption{WER across SNRs for male and female speakers, in terms of mean (solid lines) $\pm$ standard deviation (light colored areas).}
%  \label{fig:wer}
%\end{figure}

To evaluate the impact of noise on speech recognition performance across genders, we examine the WER at different SNR levels, which is displayed in the first line of Table~\ref{tab:metrics_gender_snr}.
As the SNR increases, the WER decreases for both genders, indicating improved speech recognition performance at lower noise levels. At -5~dB and 0~dB, female speech consistently exhibits a lower WER compared to male speech, suggesting that speech recognition models handle female voices slightly better under noisy conditions ($p=0.012$ and $0.048$, respectively). However, at 5~dB, where speech is more dominant over noise, the WER difference between male and female speech is not significant, confirming that gender-related effects diminish as noise interference decreases.

% P-values
% \begin{table}[t]
% \centering
% \caption{Mann-Whitney U Test Results ($p$-value) for WER Across Genders.}
% \label{tab:wer_gender}
% \begin{tabular}{l c c c}
% \hline
%  & -5~dB & 0~dB & 5~dB \\
% \hline
% Male vs. Female & \textbf{0.0000} & \textbf{0.0000} & \textbf{0.0054} \\
% \hline
% \multicolumn{4}{l}{\footnotesize{Bold indicates significant difference ($p$-value $<$ 0.05).}} \\
% \end{tabular}
% \end{table}

\subsubsection{Impact on perceptual metrics}

\begin{table}[t]
\centering
\caption{Mean Speech Recognition and Perceptual Metrics Across Genders and SNR Levels.}
\label{tab:metrics_gender_snr}
\small  % Reduce font size for better fit in IEEEtran
\begin{tabular}{l cccccc}
\hline
Metrics & \multicolumn{2}{c}{-5~dB} & \multicolumn{2}{c}{0~dB} & \multicolumn{2}{c}{5~dB} \\
\cline{2-3} \cline{4-5} \cline{6-7}
        & M        & F      & M        & F      & M        & F \\
\hline
$\text{WER}$ & 70.20   & 61.37  & 31.62    & 27.20  & 17.90    & 16.52 \\
$\text{STOI}_{in}$      & 0.46     & 0.47      & 0.57      & 0.58      & 0.69      & 0.69 \\
$\text{STOI}_{out}$     & 0.61     & 0.62      & 0.75      & 0.75      & 0.82      & 0.83 \\
$\Delta_{\text{STOI}}$   & 0.15     & 0.14      & 0.17      & 0.17      & 0.12      & 0.13 \\
$\text{PESQ}_{in}$      & 1.07     & 1.04      & 1.10      & 1.06      & 1.21      & 1.12 \\
$\text{PESQ}_{out}$     & 1.19     & 1.19      & 1.41      & 1.46      & 1.67      & 1.76 \\
$\Delta_{\text{PESQ}}$   & 0.12     & 0.14      & 0.31      & 0.39      & 0.46      & 0.64 \\
$\text{HASPI}$   & 0.38     & 0.47      & 0.84      & 0.81      & 0.96      & 0.92 \\
\hline
\end{tabular}
\end{table}

% \begin{table}[t]
% \centering
% \caption{WER across SNR level for both male and female
% }
% \label{tab:metrics_gender_sir_out}
% \small  % Reduce font size for better fit in IEEEtran
% \begin{tabular}{l cccccc}
% \hline
%  & \multicolumn{2}{c}{-5~dB} & \multicolumn{2}{c}{0~dB} & \multicolumn{2}{c}{5~dB} \\
% \cline{2-3} \cline{4-5} \cline{6-7}
%         & M        & F      & M        & F      & M        & F \\
% \hline
% Value   & 70.20   & 61.37  & 31.62    & 27.20  & 17.90    & 16.52 \\
% \hline
% \end{tabular}
% % \vspace{0.1cm}  % Add a small vertical space
% % \scriptsize  % Use a smaller font size for the note
% % \textit{Note:} M = Male, F = Female.
% \end{table}

Finally, we investigate gender-based variations in terms of perceptual metrics. The mean PESQ, STOI, and HASPI values across SNRs are presented in
Table~\ref{tab:metrics_gender_snr}. Both males and females show similar patterns in input and output STOI, with a steady increase as SNR improves, and the gap between input and output remains fairly consistent across genders. However, PESQ scores exhibit increasing disparity: female input scores are lower than males’ across all SNR levels, but their output scores tend to surpass males’ as SNR rises. STOI improvements, in contrast, remain relatively stable for both genders, whereas PESQ shows a more significant difference, with female speech exhibiting greater improvement as SNR increases. Additionaly, HASPI scores indicate that female speech tends to retain slightly higher intelligibility, particularly at lower SNR levels.

Overall, the results across metrics indicate that female speech consistently shows higher perceptual improvements (higher PESQ and HASPI scores) and lower WER at most SNR levels, suggesting a stronger benefit from SE for females. This aligns with trends observed in SIR and SAR, where female speech exhibits greater interference reduction and fewer artifacts, suggesting that the acoustic characteristics of female speech are more effectively enhanced by the SE algorithms.

\section{Conclusion}
\label{sec:conclusion}

This study highlights the need for a nuanced evaluation of multichannel SE algorithms, considering the distinct acoustic characteristics of phoneme categories and gender differences.
No utterance-level SIR differences were found between genders, but most phonemes, except affricates and sibilants, had better interference reduction in female speech, which reveals variations that are overlooked in utterance-level analysis. For artifacts, a difference was found at the utterance level and for plosives, but not for other phoneme categories.

These findings can be exploited in future work, e.g., by integrating filtering algorithms that account for phoneme-specific spectral properties into SE algorithms, or optimizing deep SE algorithms with frequency-weighted / phoneme-informed losses that prioritize spectral regions that are perceptually important. Besides, one can leverage phoneme coarticulation through the use of consonant-vowel and vowel-consonant sequences to design SE algorithms that better capture transitional dynamics and ensure a more natural speech flow. 

\bibliographystyle{IEEEbib}
\bibliography{mybib}

\begin{thebibliography}{10}

\bibitem{Esra2024}
A.~J.~S. Esra and Y.~Sukhi,
\newblock ``Optimized binaural enhancement for digital hearing aids,''
\newblock {\em Comp. Speech Lang.}, vol. 84, pp. 101554, 2024.

\bibitem{iwamoto2022bad}
K.~Iwamoto, T.~Ochiai, M.~Delcroix, R.~Ikeshita, H.~Sato, S.~Araki, and
  S.~Katagiri,
\newblock ``Impact of speech enhancement errors on {ASR},'' 2022.

\bibitem{rao2021interspeech}
W.~Rao, Y.~Fu, Y.~Hu, X.~Xu, Y.~Jv, J.~Han, Z.~Jiang, L.~Xie, Y.~Wang,
  S.~Watanabe, Z.-H. Tan, H.~Bu, T.~Yu, and S.~Shang,
\newblock ``Conferencingspeech challenge: Far-field multi-channel speech
  enhancement,''
\newblock in {\em Proc. ASRU}, 2021, pp. 679--686.

\bibitem{Benesty2008SpringerHandbook}
J.~Benesty, M.~M. Sondhi, and Y.~A. Huang, Eds.,
\newblock {\em Springer Handbook of Speech Processing},
\newblock Springer, 2008.

\bibitem{Hendriks2011}
R.~C. Hendriks and T.~Gerkmann,
\newblock ``Noise correlation matrix estimation for speech enhancement,''
\newblock {\em IEEE Trans. Audio Speech Lang. Process.}, vol. 20, no. 1, pp.
  223--233, 2012.

\bibitem{Heymann2016}
J.~Heymann, L.~Drude, and R.~Haeb-Umbach,
\newblock ``Neural network spectral mask estimation for beamforming,''
\newblock in {\em Proc. IEEE ICASSP}, 2016, pp. 196--200.

\bibitem{Nugraha2016}
A.~A. Nugraha, A.~Liutkus, and E.~Vincent,
\newblock ``Multichannel audio source separation with deep neural networks,''
\newblock {\em IEEE/ACM Trans. Audio Speech Lang. Process.}, vol. 24, no. 9,
  pp. 1652--1664, September 2016.

\bibitem{Coto-Jimenez2018}
M.~Coto-Jimenez, J.~Goddard-Close, L.~Di Persia, and H.~Leonardo Rufiner,
\newblock ``Hybrid speech enhancement with {Wiener} filters and {LSTM}
  autoencoders,''
\newblock in {\em Proc. IWOBI}, July 2018, pp. 1--8.

\bibitem{Liu2018mvdr}
Y.~Liu, A.~Ganguly, K.~Kamath, and T.~Kristjansson,
\newblock ``Neural {MVDR} beamforming with time-frequency masks,''
\newblock in {\em Proc. ICASSP}, April 2018, pp. 6717--6721.

\bibitem{Carbajal2020}
G.~Carbajal, R.~Serizel, E.~Vincent, and E.~Humbert,
\newblock ``Joint {NN}-supported multichannel echo, reverberation, and noise
  reduction,''
\newblock {\em IEEE/ACM Trans. Audio Speech Lang. Process.}, vol. 28, pp.
  2158--2173, July 2020.

\bibitem{Furnon2021}
N.~Furnon, R.~Serizel, S.~Essid, and I.~Illina,
\newblock ``{DNN}-based mask estimation for distributed speech enhancement,''
\newblock {\em IEEE/ACM Trans. Audio Speech Lang. Process.}, vol. 29, pp.
  2310--2323, June 2021.

\bibitem{wang18h_interspeech}
Z.-Q. Wang and D.~Wang,
\newblock ``All-neural multi-channel speech enhancement,''
\newblock in {\em Proc. Interspeech}, September 2018, pp. 3234--3238.

\bibitem{Luo2019}
Y.~Luo, C.~Han, N.~Mesgarani, E.~Ceolini, and S.-C. Liu,
\newblock ``Fasnet: Low-latency adaptive beamforming,''
\newblock in {\em Proc. IEEE ASRU}, December 2019, pp. 260--267.

\bibitem{Vincent2006bsseval}
E.~Vincent, R.~Gribonval, and C.~Févotte,
\newblock ``Performance measurement in blind audio source separation,''
\newblock {\em IEEE Trans. Audio Speech Lang. Process.}, vol. 14, pp.
  1462--1469, 2006.

\bibitem{LeRoux2019sisdr}
J.~Le Roux, S.~Wisdom, H.~Erdogan, and J.~R. Hershey,
\newblock ``{SDR}: Half-baked or well done?,''
\newblock in {\em Proc. IEEE ICASSP}, May 2019, pp. 626--630.

\bibitem{MillerNicely1955}
G.~A. Miller and P.~A. Nicely,
\newblock ``Perceptual confusions among english consonants,''
\newblock {\em J. Acoust. Soc. Am.}, vol. 27, pp. 338--352, 1955.

\bibitem{adachi2006intelligibility}
T.~Adachi, R.~Akahane-Yamada, and K.~Ueda,
\newblock ``Intelligibility of english phonemes in noise for native and
  non-native speakers,''
\newblock {\em Acoust. Sci. Technol.}, vol. 27, no. 5, pp. 285--289, 2006.

\bibitem{meyer2010human}
B.~T. Meyer, T.~Jürgens, T.~Wesker, T.~Brand, and B.~Kollmeier,
\newblock ``Human phoneme recognition and speech variability,''
\newblock {\em J. Acoust. Soc. Am.}, vol. 128, no. 5, pp. 3126--3141, 2010.

\bibitem{zaar2017predicting}
J.~Zaar and T.~Dau,
\newblock ``Predicting consonant recognition and confusions in normal
  hearing,''
\newblock {\em J. Acoust. Soc. Am.}, vol. 141, no. 2, pp. 1051--1064, 2017.

\bibitem{monir2024phonemescale}
N.-E. Monir, P.~Magron, and R.~Serizel,
\newblock ``A phoneme-scale assessment of multichannel speech enhancement
  algorithms,''
\newblock {\em Trends in Hearing}, vol. 28, pp. 23312165241292205, 2024.

\bibitem{Henton1986}
C.~Henton,
\newblock {\em Phonetic Sex-Specific Differences Across Languages},
\newblock Ph.D. thesis, Univ. Oxford, 1986.

\bibitem{coleman1976comparison}
R.~O. Coleman,
\newblock ``Voice quality and gender perception,''
\newblock {\em J. Speech Hear. Res.}, vol. 19, no. 1, pp. 168--180, 1976.

\bibitem{pepiot}
E.~Pépior,
\newblock ``Gender identification by voice in english and french,''
\newblock {\em Sci. Works}, vol. 49, pp. 418--430, 2011.

\bibitem{calliope1989parole}
L.~Calliope and G.~Fant,
\newblock {\em Speech and Its Automatic Processing},
\newblock Masson, 1989.

\bibitem{addadecker05_interspeech}
M.~Adda-Decker and L.~Lamel,
\newblock ``Speech recognizers and female speakers,''
\newblock in {\em Proc. Interspeech}, September 2005, pp. 2205--2208.

\bibitem{Fasnet}
Yi~Luo, Zhuo Chen, Nima Mesgarani, and Takuya Yoshioka,
\newblock ``End-to-end microphone permutation and number invariant
  multi-channel speech separation,''
\newblock in {\em Proc. IEEE ICASSP}, 2020, pp. 6394--6398.

\bibitem{Panayotov2015librispeech}
V.~Panayotov, G.~Chen, D.~Povey, and S.~Khudanpur,
\newblock ``Librispeech: {ASR} corpus from public audio books,''
\newblock in {\em Proc. IEEE ICASSP}, April 2015, pp. 5206--5210.

\bibitem{Scheibler2018}
R.~Scheibler, E.~Bezzam, and I.~Dokmanić,
\newblock ``Pyroomacoustics: Python package for audio room simulation,''
\newblock in {\em Proc. IEEE ICASSP}, 2018.

\bibitem{binaurec}
L.~Delebecque and R.~Serizel,
\newblock ``Binaurec: Dataset for binaural speech enhancement with rirs,''
\newblock in {\em Proc. EUSIPCO}, September 2023, pp. 126--130.

\bibitem{McAuliffe2017MontrealFA}
M.~McAuliffe, M.~Socolof, S.~Mihuc, M.~Wagner, and M.~Sonderegger,
\newblock ``Montreal forced aligner: Text-speech alignment with kaldi,''
\newblock in {\em Proc. Interspeech}, August 2017.

\bibitem{5483076}
A.~Bertrand and M.~Moonen,
\newblock ``Distributed adaptive signal estimation in sensor networks,''
\newblock {\em IEEE Trans. Signal Process.}, vol. 58, no. 10, pp. 5277--5291,
  2010.

\bibitem{Pariente2020Asteroid}
M.~Pariente, S.~Cornell, J.~Pons, S.~B. R., A.~Deleforge, and E.~Vincent,
\newblock ``Asteroid: {PyTorch}-based audio source separation toolkit,''
\newblock in {\em Proc. Interspeech}, 2020, pp. 2637--2641.

\bibitem{watanabe2018espnet}
S.~Watanabe, T.~Hori, S.~Karita, T.~Hayashi, J.~Nishitoba, Y.~Unno, N.~Yalta,
  J.~Heymann, M.~Wiesner, N.~Chen, A.~Renduchintala, and T.~Ochiai,
\newblock ``Espnet: End-to-end speech processing toolkit,'' 2018.

\bibitem{Rix2001}
A.~W. Rix, J.~G. Beerends, M.~P. Hollier, and A.~P. Hekstra,
\newblock ``Perceptual evaluation of speech quality ({PESQ}),''
\newblock in {\em Proc. IEEE ICASSP}, May 2001, pp. 749--752.

\bibitem{STOI}
C.~H. Taal, R.~C. Hendriks, R.~Heusdens, and J.~Jensen,
\newblock ``Intelligibility prediction of noisy speech,''
\newblock {\em IEEE Trans. Audio Speech Lang. Process.}, vol. 19, no. 7, pp.
  2125--2136, 2011.

\bibitem{haspi}
C.~Spille, B.~Kollmeier, and B.~T. Meyer,
\newblock ``Human vs. automatic speech recognition in acoustic scenes,''
\newblock {\em Comput. Speech Lang.}, vol. 52, pp. 123--140, 2018.

\bibitem{wav2vec2.0}
A.~Baevski, H.~Zhou, A.~Mohamed, and M.~Auli,
\newblock ``wav2vec 2.0: Self-supervised speech representation learning,''
  2020.

\bibitem{conformer}
A.~Gulati, J.~Qin, C.-C. Chiu, N.~Parmar, Y.~Zhang, J.~Yu, W.~Han, S.~Wang,
  Z.~Zhang, Y.~Wu, and R.~Pang,
\newblock ``Conformer: Convolution-augmented transformer for speech
  recognition,'' 2020.

\bibitem{whisper}
A.~Radford, J.~W. Kim, T.~Xu, G.~Brockman, C.~McLeavey, and I.~Sutskever,
\newblock ``Robust speech recognition via large-scale weak supervision,'' 2022.

\end{thebibliography}

\end{document}